\documentclass[12pt]{article}

\usepackage[totalheight = 23cm, totalwidth = 17cm]{geometry}
\usepackage{amssymb,amsmath,amsfonts,amsbsy,graphicx,bm}

\usepackage{color,cancel,ulem, hyperref}

\newcommand{\dis}[1]{\begin{equation}\begin{split}#1\end{split}\end{equation}}

\begin{document}

\begin{titlepage}

\begin{center}

{\LARGE \bf 
 Local gauge   invariant   operator on isometry breaking background
}

\vskip 1.0cm

{\large
Min-Seok Seo$^{a}$ 
}

\vskip 0.5cm

{\it
$^{a}$Department of Physics Education, Korea National University of Education,
\\ 
Cheongju 28173, Republic of Korea
}

\vskip 1.2cm

\end{center}

\begin{abstract}
 
 Whereas  local field operators play the crucial role in reconciling quantum mechanics  and special relativity, they are not trivially compatible with the diffeomorphism invariance of gravity.
 In order to address this issue, we consider the background geometry which breaks the isometry  spontaneously.
 Then the local gauge invariant operator can be constructed through the St\"uckelberg mechanism, where  the fluctuation of the metric in the direction of the isometry breaking combines  with that of   matter whose classical solution breaks the isometry.
 This is equivalent to introducing the clock and the rod to promote the local field operators to the gauge invariant ones.
 A typical example is  the curvature perturbation in quasi-de Sitter space arising from the spontaneous breaking of the timelike isometry.
 We also discuss the features of the local gauge invariant operator when the spacelike isometry is spontaneously broken.
 Meanwhile, even if the  local gauge invariant operators exist, it does not guarantee the reliable construction of the gauge invariant operators on the local region like the island, which is regarded as an essential ingredient to resolve the black hole information paradox.
 This is because the fluctuation of the spacetime point is accumulated in time, which in fact also gives rise to eternal inflation in quasi-de Sitter space.
 In order to suppress the fluctuation at late time, the isometry must be strongly broken by the background.
In the case of the evaporating black hole, it may be achieved by the transition to the higher dimensional black hole.

\end{abstract}

\end{titlepage}

\newpage

\section{Introduction}

 Local field operators are the main building blocks of local quantum field theory, playing a key role in   successfully reconciling  quantum mechanics with special relativity   (see, e.g., \cite{Weinberg:1995mt, Duncan:2012aja} and references therein).
Nevertheless, not all of them are physically meaningful when gravity is taken into account : for spacetime to  be   treated as a dynamical entity in a consistent way,   physical operators must be invariant under the diffeomorphism.
\footnote{See, e.g., \cite{FrancoisAndre:2023jmj, Francois:2024rdm} for recent formal treatments.}
Indeed, it has been argued that such a gauge invariant operator is typically non-local, which is realized by `dressing' the local operator at some point $x$ with a Wilson line connecting $x$ to the boundary  \cite{Donnelly:2015hta, Donnelly:2016rvo, Donnelly:2017jcd, Giddings:2018umg, Donnelly:2018nbv, Giddings:2022hba, Giddings:2025xym}.
This `dressing theorem'  can be shown by observing that if the background has a boundary or an asymptotic region,  the diffeomorphism generator $\Phi$  contains a boundary term $\Phi_{\partial}$ of   ${\cal O}(\kappa^{-1})$ where $\kappa=(8\pi G)^{1/2}$, thus can be written in the form of $\Phi=\Phi_{\rm bulk}+\Phi_{\partial}$.
The local operator $O$  in the bulk commutes with the boundary term ($[O, \Phi_{\partial}]=0$) while the commutator $[O, \Phi_{\rm bulk}]$ of   ${\cal O}(\kappa^0)$ is typically nonzero.
Then $O$ can be promoted to the gauge invariant operator by corrected by the Wilson line $\kappa O'$ connecting the bulk and the boundary, as the   commutator $[\kappa O', \Phi_{\partial}]$ of ${\cal O}(\kappa^0)$ is nonvanishing  but identical to $-[O, \Phi_{\rm bulk}]$.
In the end, the dressed  operator $O+\kappa O'$ is gauge invariant and nonlocal.

 However, if all the gauge invariant operators are nonlocal,  it becomes challenging to explain several quantum gravity phenomena.
First, it was recently suggested that the black hole information paradox can be resolved by the appearance of  a local region called the `island' in the black hole interior after the Page time \cite{Penington:2019npb, Almheiri:2019psf, Almheiri:2019hni, Penington:2019kki, Almheiri:2019qdq} (see also \cite{Almheiri:2020cfm, Raju:2020smc} for reviews).
As the operators localized on the island can be reconstructed from the operators for the Hawking radiation far away from the black hole, the entropy of the Hawking radiation does not increase indefinitely, which is consistent with unitarity.
But in the absence of local gauge invariant operator, the operators localized on the island are difficult to be well defined. 
 Second, whereas de Sitter (dS) space well describes the observed accelerating expansion of our universe \cite{WMAP:2003elm, SupernovaCosmologyProject:2003dcn}, it does not have a boundary, so the Wilson line connecting the bulk and the boundary is not defined.

 Regarding the first issue of the island, it has been suggested to construct the local operator   by explicitly breaking the diffeomorphism invariance, resulting in a massive graviton \cite{Geng:2020fxl, Geng:2020qvw, Geng:2021hlu, Geng:2023zhq, Geng:2025bcb}.
 This is motivated by the fact that when we consider the island of the anti-de Sitter (AdS) black hole   in the context of AdS/CFT correspondence \cite{Maldacena:1997re, Witten:1998qj, Gubser:1998bc},  the boundary of AdS space is assumed to be attached to the `bath' in which gravity is not dynamical and the old Hawking radiation is  contained.
  Whereas it is a convenient way to separate the Hawking radiation from gravitational dynamics, since energy is allowed to pass through the border between   AdS space and the bath, the gravitational energy-momentum tensor is not conserved.
 In other words, the diffeomorphism invariance is broken by the boundary conditions.
 This can be modelled by introducing the compensating vector field $V_\mu$ which transforms as $V_\mu\to V_\mu-  \epsilon_\mu$ under the diffeomorphism $x^\mu \to x^\mu+  \epsilon^\mu$ before being fixed to some  function of $x^\mu$.
 Then $V_\mu$ behaves like the Goldstone vector field : it combines with the metric fluctuation  $h_{\mu\nu}$ to form the would-be gauge invariant combination $\tilde{h}_{\mu\nu}=h_{\mu\nu}+\nabla_\mu V_\nu+\nabla_\nu V_\mu$, from which the graviton mass term $-(M^4/4)(\tilde{h}_{\mu\nu}\tilde{h}^{\mu\nu}-\tilde{h}^\mu_\mu \tilde{h}^\nu_\nu)$ can be written in the Lagrangian density \cite{Geng:2025bcb}. 
 This corresponds to the imitation of (the tensor version of) the St\"uckelberg (or the Higgs in a broad sense) mechanism, in the vector version of which gauge invariance is spontaneously broken as the gauge boson becomes massive by absorbing a scalar transforming nonlinearly  under the gauge transformation.
 Moreover, any local operator $O(x)$ can be promoted to the local would-be gauge invariant operator $O(x-V(x))$    which is expected to be the form of operators localized on the island   \cite{Geng:2024dbl}.
 
 Meanwhile, it was pointed out in \cite{Antonini:2025sur} that  the tension between the gauge invariance and the locality discussed so far arises only if the background has a symmetry called isometry. 
 To see this intuitively we observe that the nonlocal connection between the bulk and the boundary is imposed by the isometry.
 This may be understood  from the fact that   any   isometry breaking fluctuation raises the energy, which must be reflected in the fluctuation of the asymptotic charge like the Arnowitt-Deser-Misner (ADM) Hamiltonian.  
 In contrast, in the absence of the isometry, the asymptotic charge may remain unaffected by the fluctuation in the bulk. 
Technically, the nonexistence of the local operator on the  background having the isometry can be explained by the absence of the clock and the rod.
 To see this more concretely, suppose  the background   has a timelike isometry. 
 In this case, one finds a time direction along which the metric does not change, hence the classical solution for the matter field is independent of  the corresponding time coordinate $t$ as well. 
 Then there is no way to distinguish some particular time slice from another, which is said to be the absence of the clock.
 In the same way, the spacelike isometry prevents   distinguishing a particular spatial slice from another, then we   say that the rod does not exist.
 
  The situation changes when the background does not respect the isometry.
  As the classical solutions for the matter field as well as the metric depend on $t$,  we can distinguish different time slices defined by the values of $t$ : we now have a clock.
   This enables us to decompose any matter field, for example, the scalar field, into the $t$ dependent classical solution and the quantum fluctuation around it as $\phi(t, \mathbf{x})=\phi_0(t)+\varphi(t, \mathbf{x})$.
  If we fix the background $\phi_0(t)$,  under the time translation $t \to t+  \epsilon$ the fluctuation transforms nonlinearly  as $\varphi \to \varphi+  \epsilon \dot{\phi_0}$ where the dot indicates the derivative with respect to $t$.   
 Then the St\"uckelberg mechanism works : $\varphi$ can combine with $\zeta(t, \mathbf{x})$, a component of the metric fluctuation in the direction of the time translation (hence $\zeta\to \zeta+\epsilon$ under $t\to t+\epsilon$), to become the gauge invariant operator $\varphi-\dot{\phi}_0\zeta$, which is nothing more than the fluctuation of $\phi(t-\zeta)$.
 The fact that $t-\zeta$ is a gauge invariant combination indicates that  any local operator $O(t)$ can be promoted to the gauge invariant one $O(t-\zeta)$, which is local as well.
 \footnote{
 One may also construct the local gauge invariant operator in the form of $O(t-\varphi/\dot{\phi}_0)$, since $t-\varphi/\dot{\phi}_0$ is another gauge invariant combination.
 From the similarity of this expression to $O(x -V(x))$ where $V_\mu(x)$ is the compensating vector field mentioned above, we may interpret introducing $V_\mu(x)$ as an implement of the clock and the rod by breaking the diffeomorphism invariance, as emphasized in \cite{Geng:2024dbl}.}
 Such a breaking of timelike isometry may be an important ingredient for resolving the black hole information paradox, since the paradox compares the evaporating black hole with the initial pure state before gravitational collapse, instead of considering   eternal black hole.

 In fact,  the idea that the local gauge invariant operator can be constructed by   combining the matter and the metric fluctuations goes back to   \cite{Bardeen:1980kt} (see also discussion in \cite{Giddings:2022hba, Giddings:2025xym} which distinguishes the `gravitational dressing' to which the dressing theorem is applied from `the field dressing' where matter field also contributes to the dressing for gauge invariance).
 It was immediately used to describe the cosmological fluctuations around the Friedmann-Robertson-Walker (FRW) background which may not have a boundary, resolving the second issue \cite{Sasaki:1983kd, Brandenberger:1983vj, Halliwell:1984eu, Shirai:1987xh, Mukhanov:1988jd, Deruelle:1991sb}. 
 In particular, the curvature perturbation, the local gauge invariant operator made up of the scalar fluctuation and the trace of the spatial metric fluctuation, well describes the primordial inhomogeneity generated in the inflationary era, in which the background geometry is given by quasi-dS space  where the dS isometry is slightly broken \cite{Mukhanov:1981xt, Mukhanov:1990me}.
 The  St\"uckelberg mechanism for the curvature perturbation must be distinguished from the imitated St\"uckelberg mechanism with the compensating vector field $V_\mu$ mentioned above, since whereas the isometry is broken by the background, the diffeomorphism invariance remains unbroken.
 Indeed, since only the scalar part of the metric fluctuation is relevant, the curvature perturbation  is a separate object from the graviton, a transverse traceless tensor fluctuation of the metric which is  gauge invariant.
 In this case, the graviton remains massless.
  As the mass of the curvature perturbation is well below the inverse of the horizon radius, we can construct the effective field theory of inflation for it as well as the graviton to describe the cosmological perturbations in the inflationary era \cite{Cheung:2007st, Weinberg:2008hq}.

 In this article, we investigate  the features of local gauge invariant operators on the background that breaks the isometry spontaneously  in detail.
 \footnote{ Whereas we focus on the linear diffeomorphism  and constraints linear in fluctuations only, it was shown in \cite{Fischer:1973, Moncrief:1975xtw} (as cited in  \cite{Antonini:2025sur}) that a solution to the linearized Einstein equation around the background well approximates an exact nonlinear solution provided the variations of the constraints are surjective, i.e., the initial conditions fully determine the future Cauchy slice.
}
  For this purpose, we note that the  gauge invariance is realized by imposing constraints, from which unphysical degrees of freedom are decoupled from physical processes in a consistent way \cite{Henneaux:1994lbw}.
 As operators, these constraints play the role of the generators of the gauge transformation, the Hamiltonian and the momentum in the case of the diffeomorphism.
 Such nature of gauge invariance is well described in the Hamiltonian formalism, also known as the ADM formalism \cite{Arnowitt:1962hi}, which  has motivated the studies on   the curvature perturbation in the language of it \cite{Halliwell:1984eu, Langlois:1994ec} (see also \cite{Giesel:2017roz, Jha:2022svf} for reviews and \cite{Chowdhury:2021nxw} for a discussion in the context of AdS/CFT).
 Indeed, the ADM formalism is an essential part of  the path integral description, the optimal framework for the quantum treatment of gauge invariance \cite{Anderegg:1994xq, Prokopec:2010be, Gong:2016qpq} (see also  \cite{Binosi:2015obq, Armendariz-Picon:2016dgd} for more rigorous treatment considering the BRST symmetry).
 Our  discussion is also made based on the ADM formalism, which is reviewed in Sec. \ref{Sec:ADM}.
 In Sec. \ref{Sec:LGO}, it is shown in detail that when the isometry is spontaneously broken, how the fluctuation of the metric is combined with that of the scalar in the direction of isometry breaking to form the local gauge invariant operator.
 We first focus on the structure of the curvature perturbation on quasi-dS space where the timelike isometry is   spontaneously broken, and then move onto the case of the spontaneous breaking of the spacelike isometry.
 In both cases, while the scalar or the vector part of the metric fluctuation becomes massive by absorbing the matter fluctuation,  the graviton as a transverse  traceless fluctuation of the metric is still massless, indicating that diffeomorphism invariance is not broken. 
  Meanwhile, even though the local gauge invariant operators are defined, it does not guarantee that  gauge invariant operators supported on the local region like the island can be reliably constructed.
 It is because  the fluctuation of the spacetime point   is accumulated in time, the effect of which becomes large when the isometry breaking is too tiny.
 This another tension  between the locality and the isometry was pointed out in \cite{Antonini:2025sur}, and the same effect also can be found in quasi-dS space, resulting in eternal inflation. 
 As discussed in Sec. \ref{Sec:tension}, in the case of the evaporating black hole, even if the isometry is strongly broken, the fluctuation does not seem to be suppressed in a simple manner unless, for example, the transition to the higher dimensional black hole takes place. 
 
  We  finally note that while \cite{Folkestad:2023cze} also claimed that the locality can be compatible with diffeomorphism invariance hence the island is well defined, the meaning of the locality there differs from ours.
More concretely, the locality in \cite{Folkestad:2023cze} refers to the situation that the field configurations in two regions do not affect each other.
This can be achieved even at the classical level when the perturbation restricted to the local region solves the  linearized constraint equations  without disturbing field values in other region.
In contrast, the locality in our work is a quantum mechanical condition on operators, the commutativity of operators on different space-like separated region.
We do not need to fix the field value in some region  : the field can fluctuate in any spacetime point, which is nothing more than the quantum behavior of local operator.
In this case, the locality seems to be implemented by describing all the gravitational phenomena including the island with the local operators, but it is not enough.
 We need to see whether the accumulated fluctuation of spacetime point can be sufficiently suppressed.

\section{The ADM formalism}  
\label{Sec:ADM}
  
   We begin our discussion with the review on the  ADM formalism, where the metric in 4-dimensional spacetime is decomposed as
\dis{ds^2=-N^2 dt^2+\gamma_{ij}(N^i dt+dx^i)(N^j dt+dx^j).}
Denoting the covariant derivative with respect to $\gamma_{ij}$ by $D_i$, the extrinsic curvature is given by
\dis{K_{ij}=\frac{1}{2N}\big(\partial_t\gamma_{ij}-D_i N_j-D_j N_i\big),}
then the Einstein-Hilbert action, the pure gravity part of the action, is written as
\dis{S_{\rm EH}=\frac{1}{2\kappa^2}\int d^4x\sqrt{-g}R=\frac{1}{2\kappa^2}\int d^4 x\sqrt{\gamma}N \big(R^{(3)}+K_{ij}K^{ij}-(K^i_{~i})^2\big),}
where $\kappa^2=8\pi G$ and $R^{(3)}$ is the Ricci scalar constructed from $\gamma_{ij}$.
 From this, the canonical momentum of $\gamma_{ij}$  is obtained as
 \dis{\Pi^{ij}=\frac{1}{\sqrt{\gamma}}\frac{\delta S_{\rm EH}}{\delta \partial_t\gamma_{ij}}=\frac{1}{2\kappa^2}\big(K^{ij}-\gamma^{ij}K^k_{~k}\big).\label{eq:Piij}}
We note that due to a factor $1/\sqrt{\gamma}$, $\Pi^{ij}$ behaves as a tensor under the diffeomorphism.
 Then $S_{\rm EH}$ can be written in the first-order form,
 \dis{S_{\rm EH}=\int d^4 x \sqrt{\gamma}\big(\Pi^{ij}\partial_t\gamma_{ij}-N {\cal H}_G-N_i {\cal H}_G^i\big), }
 where
 \dis{&{\cal H}_G=2\kappa^2 \Big(\Pi_{ij}\Pi^{ij}-\frac12(\Pi^i_{~i})^2\Big)-\frac{1}{2\kappa^2}R^{(3)},
 \\
 &{\cal H}_G^i=-2D_j\Pi^{ij}.\label{eq:HG}}
 From this, one finds that the lapse $N$ and the shift $N^i$ are interpreted as Lagrange multipliers for the constraints, the Hamiltonian density ${\cal H}_G$ and the momentum density ${\cal H}_G^i$, respectively.
 
 On the other hand, in the presence of matter described by a scalar field $\phi$, the matter part of the action is given by
 \dis{S_M=\int d^4 x\sqrt{-g}\Big[-\frac12 g^{\mu\nu}\partial_\mu \phi \partial_\nu \phi-V(\phi)\Big],}
 from which the canonical momentum of $\phi$ is given by
 \dis{\Pi_\phi=\frac{1}{\sqrt{\gamma}}\frac{\delta S_{\rm EH}}{\delta \partial_t\phi}= \frac{1}{N}\partial_t\phi.}
 Then $S_M$ can be rewritten as
 \dis{S_M=\int d^4 x\sqrt{\gamma}\big(\Pi_\phi\partial_t\phi-N{\cal H}_M-N_i {\cal H}_M^i\big),}
 where the Hamiltonian and the momentum densities for matter as constraints are given by
 \dis{&{\cal H}_M=\frac12 \Pi_\phi^2+\frac12\gamma^{ij}\partial_i \phi \partial_j \phi +V(\phi),
 \\
 &{\cal H}_M^i=\Pi_\phi\partial^i\phi,}
 respectively.
 In terms of the total action $S=S_{\rm EH}+S_M$, the coefficients of $N$ and $N^i$ correspond to the total Hamiltonian density ${\cal H}={\cal H}_G+{\cal H}_M$ and the total momentum density ${\cal H}^i={\cal H}_G^i+{\cal H}_M^i$, respectively.

\section{Spontaneous breaking of isometry and local gauge invariant operator}
\label{Sec:LGO}

Now we are interested in the local gauge invariant operators made up of the matter and the metric fluctuations around the classical solutions to the equations of motion.
They can be constructed by the St\"uckelberg mechanism  when the isometry is spontaneously broken  : some component of the metric fluctuation which was unphysical when the isometry was unbroken   combines with the matter fluctuation along the direction of the broken isometry to form the physical operator which is local as well as gauge invariant.
A typical example might be the curvature perturbation in quasi-dS space,  a local gauge invariant operator arising from the spontaneous breaking of the timelike isometry.
In this section, we first revisit the construction of the curvature perturbation in light of the ADM formalism, from which the generic features of the local gauge invariant operator arising from the spontaneous breaking of isometry are discussed.
Based on this, we address the case of   the spontaneously broken spacelike isometry as well.

\subsection{Timelike isometry} 
\label{subsec:time}
 
 If the isometry of the background geometry admits the timelike Killing vector, we can take the time coordinate as an affine parameter along the direction of this Killing vector such that the metric components ${g_{(0)}}_{\mu\nu}$ are independent of time.
 For example, (A)dS space can be described by the static coordinates, where the   time coordinate $t_s$ is defined by  the timelike Killing vector   and  ${g_{(0)}}_{\mu\nu}$ are given by 
 \dis{(N_{(0)})^2=-(1-\sigma H^2r_s^2),\quad {N_{(0)}}^i=0,\quad {g_{(0)}}_{\mu\nu}dx^i dx^j=\frac{dr_s^2}{1-\sigma H^2}+r_s^2d\Omega_2^2,}
 with $d\Omega_2^2=d\theta^2+\sin^2\theta d\phi^2$.
 Here $H$ is a constant and $\sigma$ is given by $+1$ for dS and $-1$ for AdS, respectively.
 In this case, curvature tensors  
 \dis{R^{(3)}_{ijkl}=-\sigma H^2(\gamma_{il}\gamma_{jk}-\gamma_{ik}\gamma_{jl}),\quad R^{(3)}_{ij}=\sigma 2 H^2\gamma_{ij},\quad R^{(3)}=\sigma 6 H^2,}
  are consistent with the cosmological constant where  the  potential is given by the constant value, $V=3\sigma H^2/\kappa^2$,  and the classical solution for the matter field (scalar in our case) is just $\phi_0=$constant, implying that ${\Pi_{(0)}}^{ij}={\Pi_\phi}_{(0)}=0$.
In the ADM formalism, such classical solutions  can be understood as follows.
 Denoting the fluctuations around the classical solutions by $h_{ij}$ for $\gamma_{ij}$, $\varphi$ for $\phi$, $\pi^{ij}$ for $\Pi^{ij}$, and $\pi_\phi$ for $\Pi_\phi$, respectively, we can expand $S$ with respect to these fluctuations to obtain the linear order action $S_1$.
 Then equations of motion can be read off from the coefficients of the fluctuations in $S_1$.
 But since  the term $\int d^4x \sqrt{\gamma}(\Pi^{ij}\partial_t\gamma_{ij}+\Pi_\phi\partial_t \phi)$ is at most quadratic in fluctuations, it does not contribute to $S_1$ at all, so  all the equations of motion are obtained from the terms in ${\cal H}$ and ${\cal H}^i$  linear  in the fluctuations.

 The situation changes when  we do not take the time direction to be the direction of the timelike Killing vector. 
This also includes the case where  the background geometry does not admit the timelike Killing vector.
 As an example, we consider the FRW background in the flat coordinates,
 \dis{ds^2=-dt^2+a(t)^2(dr^2+r^2 d\Omega_2^2).}
 As well known, when $H(t)\equiv \dot{a}/a$ (where $\dot{a}=\partial_t a$) is a constant, or equivalently,  $a(t)=e^{Ht}$,   this background describes dS space.
 The transformation between the static and the flat coordinates are given by
 \dis{t_s=t-\frac{1}{2H}\log\big(1-H^2r^2 e^{2Ht}\big),\quad r_s=r e^{Ht},}
 and the timelike Killing vector is written as $\partial_{t_s}=\partial_t-Hr\partial_r$.
 This can be easily understood from the fact that the metric is invariant under the translation in time $t\to t+\lambda$ when it is compensated by the dilatation $r\to e^{-H\lambda}r$, or equivalently, $t_s$ translates but $r_s$ remains unchanged.
 
 Now, we include the  fluctuations of ${\cal O}(\kappa)$  to parametrize the metric and the scalar field, together with their conjugate momenta as \cite{Prokopec:2010be}
 \dis{\Pi^{ij}=\frac{P(t)}{6 \kappa^2 a(t)^4}(\delta^{ij}+\kappa \pi^{ij}(t, \mathbf{x})),\quad &\gamma_{ij}=a(t)^2(\delta_{ij}+\kappa h_{ij}(t, \mathbf{x})),
 \\
 \Pi_\phi=\frac{1}{a(t)^3}(P_\phi(t)+\pi_\phi(t, \mathbf{x})),\quad & \phi=\phi_0(t)+\varphi(t, \mathbf{x}),
 \\
 N=N_0+\kappa n(t, \mathbf{x}),\quad & N^i=\kappa n^i(t, \mathbf{x}),\label{eq:Deffluct}}
  then set $N_0=1$
 \footnote{As can be found in \eqref{eq:Piij}, $\Pi^{ij}$ contains an overall factor $\kappa^{-2}$, which is reflected in the parametrization as well.}.
 Then the zeroth order action is given by
 \dis{S_0=\int d^4 x\Big[\frac{P}{\kappa^2}\dot{a}+P_\phi \dot{ \phi_0}-N_0\Big(- \frac{P^2}{12 \kappa^2 a}+\frac{P_\phi^2}{2a^3}+a^3V(\phi)\Big)\Big].}
 Varying this with respect to the classical fields, one finds that the following equations of motion are satisfied : 
 \dis{\frac{\delta S_0}{\delta P}=0 &:\quad\quad \dot{ a}=- \frac{P}{6a},
 \\
 \frac{\delta S_0}{\delta a}=0 &:\quad\quad \dot{P} =-\frac{P^2}{12a^2}+\frac32 \kappa^2\frac{P_\phi^2}{a^4}-3 \kappa^2 a^2 V,
 \\
 \frac{\delta S_0}{\delta P_\phi}=0 &:\quad\quad \dot{ \phi_0}=\frac{P_\phi}{a^3},
 \\
 \frac{\delta S_0}{\delta \phi}=0 &:\quad\quad \dot{P_\phi} = -a^3 \frac{dV}{d\phi_0},
 \\
  \frac{\delta S_0}{\delta N_0}=0 &:\quad\quad \frac{P^2}{12\kappa^2 a }  = \frac12 \frac{P_\phi^2}{a^3}+a^3 V.\label{eq:EoM}}
  They are equivalent to well known relations in cosmology, 
  \dis{&H^2=\frac{\kappa^2}{3}\Big(\frac12 \dot{\phi_0}^2 +V\Big),\quad\quad \dot{H}=-\frac12\kappa^2\dot{\phi_0}^2
  \\
  &\ddot{\phi_0}+3 H \dot{\phi_0} +\frac{dV}{d\phi_0}=0.}
 From them one immediately notices that only when $\dot{\phi_0}=0$ the background becomes dS space, i.e., $\dot{H}=0$ and $dV/d\phi_0=0$, thus the timelike Killing vector exists.
For this reason, $\dot{\phi_0}$ is interpreted as an order parameter for the spontaneous breaking of timelike isometry.
More precisely, one can define the dimensionless order parameter known as the `slow-roll parameter',
\dis{\epsilon_H=\frac12 \kappa^2\frac{\dot{\phi_0}^2}{H^2}=-\frac{\dot{H}}{H^2}.\label{eq:epsilonH}}
Here the last expression   contains the geometric information, showing that $\epsilon_H$ is nonvanishing only if the timelike isometry is spontaneously broken by the background, hence $H$ is no longer  a constant. 
Since $H$ in (quasi-)dS space is interpreted as the inverse of the horizon radius $R_{\rm hor}$, it can be rewritten as 
\dis{\epsilon_H=\dot{R}_{\rm hor}.}

 The role of the order parameter $\dot{\phi_0}$ (hence $\epsilon_H$) in constructing the local gauge invariant operator can be clearly understood from the  structure of the linear order action.
 For the later discussion, we divide the linear order action into the Hamiltonian part ( $-\int dt H_{1, G}$ for the gravitational and $-\int dt H_{1, M}$ for the  matter part)  and the rest $S_{1, d}$ :
 \dis{S_{1, d}&=\int d^4x \sqrt{\gamma}(\Pi^{ij}\dot{\gamma}_{ij}+\Pi_\phi\dot{\phi})\Big|_{\rm 1st}
 \\
 &=\int d^4 x\Big[ \frac{P}{3 \kappa}(\dot{ a} )\pi+\frac{a}{6 \kappa}(-\dot{P}+H P)h+(\dot{\phi_0})\pi_\phi -(\dot{P_\phi})\varphi+\frac{\kappa a}{6}\Big(\frac{3P_\phi}{a}\dot{ \phi_0}\Big)h\Big], }
 \dis{-\int dt H_{1, G}&=-\int d^4x \sqrt{\gamma}N{\cal H}_G\Big|_{\rm 1st} 
 \\
 &= \int d^4x  \Big[\frac{P}{3 \kappa}\Big( \frac{P}{6 a}\Big)\pi+ \frac{ a}{6\kappa}\Big(\frac{P^2}{12   a^2}\Big)h+\kappa n \Big(\frac{P^2}{12\kappa^2 a}\Big)+ \frac{a^2}{2\kappa }\nabla_i(\nabla_j h^{ij}-\nabla^i h)\Big],}
 \dis{-\int dt &H_{1, M}=-\int d^4x \sqrt{\gamma}N{\cal H}_M\Big|_{\rm 1st} 
 \\
 &= \int d^4x \Big[\frac{\kappa a}{6}\Big(-\frac32 \frac{P_\phi^2}{a^4}-3a^2 V\Big)h+\Big(-\frac{P_\phi}{a^3}\Big)\pi_\phi-\Big(a^3\frac{dV}{d\phi_0}\Big)\varphi+ \kappa n\Big(-\frac{P_\phi^2}{2a^3}-a^3V\Big)\Big],
 }
where indices  are raised and lowered by $\delta^{ij}$ and $\delta_{ij}$, respectively (for example, $h$ indicates $\delta^{ij}h_{ij}$).
 We note that the last surface term  in $\int dt H_{1,G}$ originates from the 3-dimensional Ricci tensor term $-[1/(2\kappa^2)]R^{(3)}$ in \eqref{eq:HG}.
 It is of ${\cal O}(\kappa^{-1})$ since we take a term linear in $\kappa h_{ij}$ of the Ricci tensor term which contains a factor $\kappa^{-2}$.
 As commented in the Introduction, if the background geometry has the boundary, this surface term may not vanish, then for gauge invariance, the  local operator which transforms nontrivially under the diffeomorphism can be dressed by the Wilson line, the nonlocal operator of ${\cal O}(\kappa)$ which does not vanish at the boundary \cite{Donnelly:2016rvo} (see also discussion in \cite{Geng:2021hlu}).
 However, (quasi-)dS space we are considering does not have a boundary, so the surface term vanishes.
 Then we cannot construct the gauge invariant operator by the nonlocal dressing.
Instead, as the dS isometry is spontaneously broken,    $\dot{\phi_0}$ (hence $\dot{H}$) does not vanish, and this leads to another way to construct the gauge invariant operator  which is now local.
 
 To see this, we infer from the expression for the linear order action that unlike the previous case where the timelike Killing vector exists and defines the time direction,  each coefficient of the fluctuations ($\pi$, $h$, $\pi_\phi$, $\varphi$) except for $n$  in the linear order Hamiltonian $H_{1, G}+H_{1, M}$ does not vanish.
 \footnote{Of course, the total linear order Hamiltonian $H_{1, G}+H_{1, M}$ vanishes since it is a constraint, providing the relations between fluctuations. }
 Since the total linear order action $S_{1, d}-\int dt (H_{1, G}+H_{1, M})$   is equivalent to the variation of the action around the classical solutions to the equations of motion, it must vanish, but this is achieved by the cancellation between coefficient in $S_{1, d}$ and that in the Hamiltonian part for each fluctuation.
 For instance, the coefficient of $[a/(6\kappa)]h$ in $S_{1, d}$ given by  $-\dot{P}+H P+(3\kappa^2 P_\phi/a)\dot{\phi_0}$ and the corresponding one in the Hamiltonian part given by  $[P^2/(12 a^2)]-\kappa^2[3P_\phi^2/(2a^4)]-3\kappa^2 a^2 V$ are added to vanish, which is consistent with the equations of motion given by \eqref{eq:EoM}.
 In the same way, the coefficients of $\pi$ and $\pi_\phi$  in the Hamiltonian part are nothing more than the minus of the corresponding coefficients in $S_{1, d}$ :
 \dis{&-\int d^3 x \frac{P}{3\kappa}\Big( \frac{P}{6a}\Big)\pi= \int d^3 x \frac{P}{3\kappa}(\dot{a})\pi,
 \\
 & \int d^3 x \Big(\frac{P_\phi}{a^3}\Big)\pi_\phi= \int d^3 x (\dot{ \phi_0})\pi_\phi. }
 Now, the commutation relations between the fluctuations and their canonical momenta can be found from  following Poisson brackets,
 \dis{&[\gamma_{ij}(t, \mathbf{x}), \Pi^{kl}(t, \mathbf{y})]_{\rm PB}=\frac12(\delta_i^k\delta_j^l+\delta_i^l\delta_j^k)\frac{1}{\sqrt{\gamma}}\delta^{(3)}(\mathbf{x}-\mathbf{y}),
 \\
 &[\phi(t, \mathbf{x}), \Pi_\phi(t, \mathbf{y})]_{\rm PB}=\frac{1}{\sqrt{\gamma}}\delta^{(3)}(\mathbf{x}-\mathbf{y}),\label{eq:PoissonO}}
 which are equivalent to (see \eqref{eq:Deffluct})
  \dis{&[h_{ij}(t, \mathbf{x}), \pi^{kl}(t, \mathbf{y})]_{\rm PB}=\frac12(\delta_i^k\delta_j^l+\delta_i^l\delta_j^k)\frac{6}{P(t)a(t)}\delta^{(3)}(\mathbf{x}-\mathbf{y}),
 \\
 &[\varphi(t, \mathbf{x}), \pi_\phi(t, \mathbf{y})]_{\rm PB}=\delta^{(3)}(\mathbf{x}-\mathbf{y}).}
Then we obtain
 \dis{&\delta h_{ij}= [h_{ij}(t, \mathbf{x}), \int d^3x \sqrt{\gamma}N{\cal H}_G\Big|_{\rm 1st} \epsilon(t, \mathbf{x}) ]_{\rm PB}=\delta_{ij}\frac{2H}{\kappa}\epsilon,
 \\ 
 &\delta\varphi=[\varphi(t, \mathbf{x}), \int d^3x \sqrt{\gamma}N{\cal H}_G\Big|_{\rm 1st} \epsilon(t, \mathbf{x}) ]_{\rm PB}]_{\rm PB}=\dot{\phi_0}\epsilon.}
Since  the Hamiltonian generates the time translation, the diagonal part of the spatial metric fluctuation $\zeta$  defined by  $h_{ij}=2\zeta\delta_{ij}+\cdots$  and the scalar fluctuation $\varphi$ transform as $\zeta \to \zeta+ (H/\kappa)\epsilon$ and $\varphi \to \varphi+\dot{\phi_0} \epsilon$, respectively,  under $t \to t+\epsilon$.

 Then we can construct the local gauge invariant operator  as follows.
 Since  $\zeta$ transforms in a nontrivial way under the time translation it is not gauge invariant. 
  This is consistent with the fact that among the metric fluctuations, only the transverse  traceless one $h_{ij}^{TT}$ is invariant under the linear diffeomorphism, hence physical.
  \footnote{This can be shown by noting that $h_{ij}^{TT}$ can be expressed in terms of the projection $P_{ij}=\delta_{ij}-n_in_j$ with $n_i=k_i/(k_jk^j)$ in momentum space as (see, e.g., Box 35.1 of \cite{Misner:1973prb})
  \dis{h_{ij}^{TT}=P_i^{~l}P_j^{~m}h_{lm}-\frac12 P_{ij}(P^{lm}h_{lm}).}
  Since $P_{ij}k^j=0$, one finds that $\delta h_{ij}^{TT}=0$ under $h_{ij}\to h_{ij}+\partial_i \xi_j+\partial_j \xi_i$.\label{foot:TT}}
However, when  the timelike dS isometry is broken by nonzero $\dot{\phi_0}$,   $\varphi$ as well as $\zeta$ transform nonlinearly under the time translation, from which we can construct the local  gauge invariant operator ${\cal R}$ defined by 
\dis{{\cal R}=\zeta-\frac{H}{\kappa \dot{\phi_0}}\varphi =\zeta-\frac{1}{\sqrt{2 \epsilon_H}}\varphi. \label{eq:R}}
This is known as the  curvature perturbation (or the Mukhanov-Sasaki variable)  \cite{Anderegg:1994xq, Mukhanov:1985rz, Sasaki:1986hm}, and the process described so far is a kind of  the St\"uckelberg   mechanism.
 Indeed, the quadratic part of the action is given in terms of ${\cal R}$ and $h_{ij}^{TT}$ only \cite{Prokopec:2010be, Gong:2016qpq}  :
 \dis{S_2=\int d^4 x a^3\Big[2\epsilon_H \frac12\Big((\partial_t {\cal R})^2-\frac{1}{a^2}(\partial_i {\cal R})^2\Big)+\frac{1}{8\kappa^2}\Big((\partial_t h_{lm}^{TT})^2-\frac{1}{a^2}(\partial_i h_{lm}^{TT})^2 \Big)\Big].}

 We can also interpret the curvature perturbation  as the promotion of the scalar field $\phi(t, \mathbf{x})$ which is local but not gauge invariant to the local gauge invariant operator.
  That is, since the fluctuation $\zeta$ transforms as $\zeta \to \zeta+ (H/\kappa)\epsilon^0 \delta^\mu_0$ under the diffeomorphism $x^\mu \to x^\mu+\epsilon^\mu$, the scalar field can be promoted to the gauge invariant operator when we replace $t$ by  the gauge invariant combination $t-(\kappa/H)\zeta$ as (see also \cite{Geng:2024dbl})
 \dis{\phi\Big(t-\kappa\frac{\zeta}{H}, \mathbf{x}\Big)\simeq \phi(t, \mathbf{x})-\kappa\dot{\phi_0}\frac{\zeta(t, \mathbf{x})}{H}=\phi_0(t)+\varphi(t, \mathbf{x})-\kappa\dot{\phi_0}\frac{\zeta(t, \mathbf{x})}{H}=\phi_0(t)-\sqrt{2\epsilon_H}{\cal R}(t, \mathbf{x}),}
 where we use the fact that $\zeta$ is a small fluctuation.
  Since any scalar operator $O$ transforms in the same way as $\phi$ under the diffeomorphism, we can construct the generic local gauge invariant operator given by the combination
  \dis{O\Big(t-\kappa\frac{\zeta}{H}, \mathbf{x}\Big)-O_0(t)= o(t, \mathbf{x})-\kappa\frac{\dot{O_0}}{H}\zeta,}
   where $O_0(t)$ is the classical solution and $o(t, \mathbf{x})$ is the fluctuation of $O$ around $O_0$.
   As an example, we can consider the case where  the timelike isometry is broken by the classical solutions of multiple scalar fields $\phi^a$ ($a=1, 2, \cdots , N$) \cite{Gong:2016qpq} (see also \cite{Achucarro:2012sm}).
   Then  we can decompose the fluctuation of the scalar field $\varphi^a$   into the components in the direction perpendicular and parallel   to $\dot{\phi}^a_0$ denoted by $\varphi_\perp(t, \mathbf{x})$ and $\kappa \dot{\phi}^a_0\pi(t, \mathbf{x})$, respectively, such that two terms of the decomposition 
   \dis{\phi^a\Big(t-\kappa\frac{\zeta}{H}, \mathbf{x}\Big)=\varphi^a_\perp(t, \mathbf{x})- \kappa\frac{\dot{\phi}^a_0}{H}\big(\zeta- H\pi\big)  (t, \mathbf{x})}
  are gauge invariant respectively.
  
  Finally, we note that since the spontaneously broken dS isometry is a combination of the shift of $t$ and the rescaling of $r$, the fluctuation $\varphi$ can be interpreted as the Goldstone boson  associated with  the dilatation of the spatial coordinates, as well as that associated with the time translation.
 Moreover, another dS isometry, the special conformal invariance with three generators is also spontaneously broken \cite{Gong:2017wgx},  but unlike quantum field theory in Minkowski space,  the number of  Goldstone bosons associated with the spontaneously broken symmetries is just one, rather than four.
  In fact, for a generic background, the number of Goldstone modes does not always  coincide with the number of broken symmetries \cite{Low:2001bw, Hidaka:2014fra}.

\subsection{Spacelike isometry}
\label{subsec:space}

 We now discuss construction of the local gauge invariant operator when the spacelike isometry is spontaneously broken.
 For ease of discussion, we focus on the case where the isometry associated with the translation in some specific direction, say, $x$ direction, is spontaneously broken by the background.
  This can be realized when the background metric components and the classical solution of the matter field depend on $x$ exclusively, i.e., independent of $y$ and $z$.
 Then the rotation isometry  is spontaneously broken as well, since spacetime everywhere isotropic is homogeneous (see, e.g., Ch. 27.3 of \cite{Misner:1973prb}).
To proceed, as we did in the case of the timelike isometry, we decompose the metric and the scalar field into the solutions to the equations of motion and fluctuations around them,
   \dis{\Pi^{ij}=\frac{1}{\kappa^2}\big(\Pi_{(0)}^{ij}(x)+\kappa\pi^{ij}(t, \mathbf{x})\big),\quad &\gamma_{ij}= {\gamma_{(0)}}_{ij}(x)+\kappa h_{ij}(t, \mathbf{x}),
 \\
 \Pi_\phi=P_\phi(x)+\pi_\phi(t,\mathbf{x}),\quad & \phi=\phi_0(x)+\varphi(t, \mathbf{x}).}
 If the spatial metric is multiplied by the overall scale factor $a(t)^2$, we divide it by $a(t)^2$ to obtain $\gamma_{ij}$ above.
 
  Meanwhile, the spatial translation in the direction of the spacelike vector $\xi$ is generated by the momentum $\int d^3x\sqrt{\gamma} ({\cal H}_G^i+{\cal H}_M^i)\xi_i$.
 To see this, we infer from \eqref{eq:PoissonO} that the Poisson brackets for the fluctuations are given by
   \dis{&[h_{ij}(t, \mathbf{x}), \pi^{kl}(t, \mathbf{y})]_{\rm PB}=\frac12(\delta_i^k\delta_j^l+\delta_i^l\delta_j^k)\frac{1}{\sqrt{\gamma}}\delta^{(3)}(\mathbf{x}-\mathbf{y}),
 \\
 &[\varphi(t, \mathbf{x}), \pi_\phi(t, \mathbf{y})]_{\rm PB}=\frac{1}{\sqrt{\gamma}}\delta^{(3)}(\mathbf{x}-\mathbf{y}),}
 from which we obtain
 {\small
  \dis{&\delta h_{ij}=[h_{ij}, \int d^3x \sqrt{\gamma}{\cal H}_G^k \Big|_{\rm 1st}  \xi_k]_{\rm PB}=[h_{ij}, \int d^3x \sqrt{\gamma} \frac{1}{\kappa^2}(\Pi_{(0)}^{kl}+\kappa\pi^{kl})2D_k \xi_l]_{\rm PB}=\frac{1}{\kappa}(D_i \xi_j+D_j\xi_i),
  \\
  &\delta\varphi=[\varphi, \int d^3x \sqrt{\gamma}{\cal H}_G^k \Big|_{\rm 1st} \Big|_{\rm 1st}  \xi_k]_{\rm PB}=[\varphi, \int d^3x  \sqrt{\gamma}(P_\phi+\pi_\phi)\partial^k \phi \xi_k]_{\rm PB}=\partial^k \phi_0\xi_k,\label{eq:transPoisson}}   } 
as expected.
 We also note that
 \dis{\delta \partial_i \varphi=[\partial_i\varphi, \int d^3x \sqrt{\gamma}{\cal H}_G^k\Big|_{\rm 1st} \xi_k]_{\rm PB}=\xi^k \partial_k\partial_i \phi_0 +\partial_k \phi_0\partial_i \xi^k.}
Replacing $\partial_i \varphi$ by the fluctuation $v_i$ of the vector field $V_i={V_{(0)}}_i+v_i$, it becomes $\delta v_i=\xi^k  \partial_k {V_{(0)}}_i+{V_{(0)}}_k \partial_i \xi^k$,  which is nothing more than the diffeomorphism   transformation of the vector field.

 Then under the translation in the $x$ direction,  $\xi^i =\epsilon(t, \mathbf{x}) \delta_x^i $, or equivalently,  $\xi_i = {\gamma_{(0)}}_{ix}\epsilon(t, \mathbf{x})$,  the fluctuation $\delta h_{ij}$ is given by
 \dis{\kappa\delta h_{ij}= D_i \xi_j+D_j\xi_i =  {\gamma_{(0)}}_{jx}\partial_i \epsilon+{\gamma_{(0)}}_{ix}\partial_j \epsilon+\big[\delta_i^x  {\gamma_{(0)}}'_{jx}+\delta_j^x  {\gamma_{(0)}}'_{ix}-2\Gamma^k_{~ij}{\gamma_{(0)}}_{kx}\big]\epsilon,\label{eq:delhij}}
where ${\gamma_{(0)}}'_{ij}$ denotes   $\partial_x {\gamma_{(0)}}_{ij}$.
 Since the Christoffel symbols   are given by
 \dis{\Gamma^k_{~ij}=\frac12\gamma_{(0)}^{kl}\big(\delta_i^x  {\gamma_{(0)}}'_{lj}+\delta_j^x  {\gamma_{(0)}}'_{li}-\delta^{x}_l  {\gamma_{(0)}}'_{ij}\big),}
 we obtain
 \dis{\kappa\delta h_{ij}= {\gamma_{(0)}}_{jx}\partial_i \epsilon+{\gamma_{(0)}}_{ix}\partial_j \epsilon+ {\gamma_{(0)}}'_{ij}\epsilon.}
 That is,  unlike the case of the timelike isometry,  $\delta h_{ij}$ contains not only the shift  ${\gamma_0}'_{ij}\epsilon$ but also the $\partial_i \epsilon$ term except for the unrealistic case of ${\gamma_{(0)}}_{ix}=0$.
Comparing with the transformations of the vector and the scalar fields whose classical solutions depend on $x$ only,
 \dis{\delta v_i=\epsilon  {V_{(0)}}'_i+{V_{(0)}}_x   \partial_i \epsilon,\quad\quad 
 \varphi=\phi'_0 \epsilon,}
  one finds the local gauge invariant operator 
\dis{\kappa h_{ij}-\Big(\frac{{\gamma_{(0)}}_{ix}}{{V_{(0)}}_x}v_j+\frac{{\gamma_{(0)}}_{jx}}{{V_{(0)}}_x}v_i \Big)+\frac{1}{\phi'_0}\Big({\gamma_{(0)}}_{ix}\frac{{V_{(0)}}'_j}{{V_{(0)}}_x}+{\gamma_{(0)}}_{jx}\frac{{V_{(0)}}'_i}{{V_{(0)}}_x} -{\gamma_{(0)}}'_{ij}  \Big)\varphi,\label{eq:trial}}
where the $\partial_i \epsilon$ term in $\delta h_{ij}$ is cancelled by that in $\delta v_i$. 
Of course, we may replace ${V_{(0)}}_{ x}$ and $v_i$ by $\phi'_0$ and $\partial_i\varphi$, respectively.

  At first glance, this seems to indicate that the graviton becomes massive by absorbing the vector and the scalar fluctuations.
  However, the background breaks the isometry  but not diffeomorphism, so the graviton should be massless according to the equivalence principle.
  Moreover, in the gravity-scalar system where the vector does not have a nontrivial classical solution, i.e.,  $V_{(0)}=0$, the role of $v_i$ is played by $\partial_i\varphi$,   then  the tensor fluctuation $h_{ij}$ absorbs only one more degree of freedom $\varphi$.
 This does not realize  the massive graviton having $5$ degrees of freedom.
  
   In fact, this confusion can be resolved by noticing that the transverse tracelss tensor fluctuation $h_{ij}^{TT}$ which is identified with the graviton is a gauge invariant operator.
  This is true in the local flat space (see footnote \ref{foot:TT}), and also can be found by comparing the covariant decomposition of the metric \cite{York:1974psa}
\dis{h_{ij}=h_{ij}^{TT}+\Big(D_i h_j+D_j h_i-\frac23 {\gamma_{(0)}}_{ij} D_k h^k\Big)+\frac13 {\gamma_{(0)}}_{ij} h, }
 with 
\dis{\kappa \delta h_{ij}=D_i \xi_j+D_j \xi_i=\Big(D_i \xi_j+D_j \xi_i-\frac23 {\gamma_{(0)}}_{ij} D_k \xi^k \Big)+\frac23{\gamma_{(0)}}_{ij}D_k \xi^k,  }
which explicitly shows that
\dis{\kappa \delta h_i =\xi_i={\gamma_{(0)}}_{ix} \epsilon,\quad\quad \kappa \delta h=2 D_i \xi^i=2\partial_x \epsilon +{\gamma_{(0)}}^{ij}{\gamma_{(0)}}'_{ij}\epsilon. \label{eq:varh} }     
Indeed, the simple example ${\gamma_{(0)}}_{ij}=a(x)^2 \delta_{ij}$ gives $\kappa \delta h=2\partial_x \epsilon+2(a'/a)\epsilon$, which resembles $\kappa \delta h=6 H\epsilon$ in the case of the spontaneous breaking of the timelike isometry discussed in Sec. \ref{subsec:time}, except for the additional $\partial_x\epsilon$ term.

 We note that $\delta h=0$ is obtained for the specific choice of  $\xi^i=\epsilon\delta^i_x$ satisfying $D_i \xi^i=0$.
 Using the Jacobi's formula for an arbitrary matrix $M$, Tr$(M^{-1}\delta M)=\delta{\rm det}M/{\rm det}M$, one finds that $D_i\xi^i$ given by the second equation in \eqref{eq:varh} vanishes provided
 \dis{\epsilon(\mathbf{x})=\frac{\epsilon_0(y, z)}{\sqrt{\gamma (x)}}.}
In the example of ${\gamma_{(0)}}_{ij}=a(x)^2 \delta_{ij}$, this corresponds to $\epsilon=\epsilon_0(y, z)/a(x)^3$, giving
\dis{D^i(D_i\xi_j+D_j\xi_i)=\Big[\partial_k\partial_k\epsilon +5\frac{a'}{a}\partial_x \epsilon+3\Big(\frac{a'}{a}\Big)^2\epsilon+3\frac{a''}{a}\epsilon\Big]\delta_{jx}+\partial_x\partial_j\epsilon+3\frac{a'}{a}\partial_j\epsilon,}
or equivalently,
\dis{&D^i(D_i\xi_x+D_x\xi_i)=-9\Big(\frac{a'}{a}\Big)^2\frac{\epsilon_0}{a^3}-3\frac{a''}{a}\frac{\epsilon_0}{a^3}+\frac{\partial_y \epsilon_0+\partial_z\epsilon_0}{a^3},
\\
&D^i(D_i\xi_j+D_j\xi_i)=0\quad\quad (j \ne x).}
Whereas $D^i(D_i\xi_j+D_j\xi_i)=0$ for $a=a_0(x-x_0)^{1/4}$ and $\epsilon_0=$ constant, this is unphysical since $x$ extends from $-\infty$ to $\infty$.
This shows that the variation $D_i\xi_j+D_j\xi_i$ cannot be traceless and transverse simultaneously, hence cannot be the variation of $h_{ij}^{TT}$, as expected.

 From \eqref{eq:varh}, we can construct the local gauge invariant operators,
 \dis{&{\cal V}_i=h_i -\frac{{\gamma_{(0)}}_{ix}}{\kappa \phi'_0}\varphi,
 \\
 &{\cal S}=h-\frac{2}{\kappa {V_{(0)}}_x}v_x+\frac{1}{\kappa\phi'_0}\Big(\frac{2{V_{(0)}}'_x}{{V_{(0)}}_x}-{\gamma_{(0)}^{ij}{\gamma_{(0)}}'_{ij}}\Big)\varphi.}
 The first operator ${\cal V}_i$  can be interpreted as a massive gauge boson arising from the St\"uckelberg mechanism : the vector part of the metric fluctuation combines with the scalar fluctuation to become a physical massive gauge boson.
 The second operator ${\cal S}$ is   a combination of the vector fluctuation and the scalar ones,  $h$ and $\varphi$, but in this case, only one component of the vector field along the isometry breaking direction ($x$ direction in our case) is relevant.
 This is because even after the isometry breaking,  translation and rotation on the $y-z$ plane still remain as isometries, and $v_x$ is a scalar with respect to SO(2) rotation on the $y-z$ plane.
 In the gravity-scalar system where the vector field is irrelevant, ${V_{(0)}}_x$ and $v_x$ are replaced by $\phi'_0$ and $\partial_i\varphi$, respectively.
 We also note that ${\cal S}$ can be obtained by contracting  \eqref{eq:trial} with ${\gamma_{(0)}^{ij}}$.

 As in the case of the spontaneous breaking of the timelike isometry, we can   promote any local scalar operator to be gauge invariant  using the fact that the combination $x-  \kappa {\gamma_{(0)}^{ix}}h_i$ is gauge invariant :
 \dis{O(t, x-  \kappa {\gamma_{(0)}^{ix}}h_i , y, z)-O_0(x) \simeq   o(t, \mathbf{x})-\kappa{O_0}'  {\gamma_{(0)}^{ix}}h_i.}
 When $O$ is given by a scalar field $\phi$, it is nothing more than $-\kappa{\phi_0}'  {\gamma_{(0)}^{ix}}{\cal V}_i$.
 In the most general case where the spacelike isometry is completely broken, that is, the translation isometries in all three spatial directions are broken, the gauge invariant combination would be
 \dis{x^i-\kappa {\gamma_{(0)}^{ji}}h_j,}
 which can be confirmed from the transformation $\kappa \delta h_i=\xi_i={\gamma_{(0)}}_{ij} \epsilon^j$ under $x^i \to x^i+\epsilon^i$.
  We can also construct the local gauge invariant operator using another gauge invariant combination 
\dis{x+\frac{1}{\frac{2{V_{(0)}}'_x}{{V_{(0)}}_x}-{\gamma_{(0)}^{ij}{\gamma_{(0)}}'_{ij}}}\Big(\kappa h -\frac{2}{{V_{(0)}}_x}v_x\Big)\label{eq:xprom}} 
which is relevant to ${\cal S}$.
 In this case, however, the problem arises   when both the timelike and the spacelike isometries are spontaneously broken, since the diffeomorphism transformation of $h$ has to cancel those of $t$ and $x$ simultaneously.
 To see this, suppose we try to promote a local field to the gauge invariant one by replacing $t$ and $x$ by $t-(\kappa/H)\zeta=t-6(\kappa/H)h+\cdots$ and \eqref{eq:xprom}, respectively.
 Then under the time translation $t\to t+\epsilon_0$, $h$ also shifts to make $t-(\kappa/H)\zeta$  invariant, but it also leads to the appearance of $\epsilon_0$   in \eqref{eq:xprom}.
  Additional attempt to cancel this $\epsilon_0$ dependence by the translation in the $x$ direction (under which  $h$ as well as $v_x$ transform) again moves   $\epsilon_0$ to the former combination, $t-(\kappa/H)\zeta$. 
  This suggests that if more than one isometries are spontaneously broken, the gauge invariance must be achieved by promoting different components of the metric ($h$ and $h_i$ in our case) physical.

 \section{Accumulated fluctuation of the spacetime point}
 \label{Sec:tension}

 As mentioned in Introduction, recent studies suggest that the information paradox can be resolved if one can construct the gauge invariant operators supported exclusively on the local
  region called island, and these operators describe the Hawking radiation far away from the event horizon of black hole.
At first glance, it seems that in the presence of the local gauge invariant operator $O(x)$ arising from the spontaneous breaking of the isometry, we can always construct the gauge invariant operator supported on the island   given by
\dis{O_{\rm island}=\int d^4 x f(x) O(x),}
where $f(x)$ is a smooth compactly supported function on the island, without breaking the diffeomorphism invariance  \cite{Antonini:2025sur}.
However, it was pointed out in Sec. 4.6 of  \cite{Antonini:2025sur} that there are several subtleties in such construction. 
As will be discussed in this section, some of them also appear in the inflationary cosmology where the background is given by quasi-dS space.

The first issue is that, after the Page time   when the island begins to appear, the uncertainties in the geometry at time $t$ increases as $t^{1/2}$, which may push the operator $O_{\rm island}$ outside the island.
This happens when   the uncertainties become  too large that   the coordinate labelling the point in the island may not be $x^\mu$ at which the value of $f(x)$ does not vanish, but  $x^\mu+\delta x^\mu$ at which $f(x+\delta x)=0$.
Such uncertainties also can be found in quasi-dS space.
 To see this, we recall that as the universe expands almost exponentially, the Fourier mode of the curvature perturbation ${\cal R}_{\mathbf k}$ with the wavenumber $k$   is frozen after $t=H^{-1}\log(k/H)$ (see footnote \ref{foot:tk}) and behaves as the classical fluctuation : the quantum interference effects are suppressed \cite{Burgess:2006jn, Burgess:2014eoa, Nelson:2016kjm, Shandera:2017qkg, Martin:2018zbe, Gong:2019yyz}.
 Since   ${\cal R}$ contains the fluctuation of the scalar $\varphi$, this frozen mode can be treated as the fluctuation of the classical trajectory $\phi_0(t)$ scaling as $t^{1/2}$ \cite{Vilenkin:1982wt, Linde:1982uu, Starobinsky:1982ee}, 
 \dis{\sqrt{\delta \big(\phi_0(t)^2 \big)}= \Big(\frac{H}{2\pi}\Big) (H \Delta t)^{1/2}. }
Moreover, the fact that $\phi_0(t)$ plays the role of the clock implies that this fluctuation is interpreted as the fluctuation in time.
Another way to see this is to notice that the metric fluctuation $\gamma_{ij}=a(t)^2 (\delta_{ij}+2\kappa\zeta \delta_{ij})$ can be regarded as  the leading term of $a^2 e^{2\kappa\zeta}\delta_{ij}$.
 Since $a=e^{Ht}$ for dS space, for a sufficiently small value of $\epsilon_H$, the fluctuation in $\zeta$ (which is promoted to the gauge invariant operator ${\cal R}$ by \eqref{eq:R})  is interpreted as  the fluctuation in time  given by
 \footnote{Here we need to distinguish $\delta t$ from $\Delta t$.
 While  $\Delta t$ is the interval of time $t$ determined by the classical solution $\phi_0(t)$, $\delta t$ given by \eqref{eq:tfluc} is the fluctuation in the value of $t$. 
 We also note that $t$ also can be measured by the moment when the mode of the wavenumber $k$ is frozen, satisfying $k = a(t) H$.
  Since $a(t)\simeq e^{Ht}$, $t$ and $k$ are related by $t=H^{-1}\log(k/H)$.
 Then $\Delta t$ as the interval between two moments $t_i$ and $t_f$ can be written as $\log(k_f/k_i)$ where $k_i=a(t_i)H$ and $k_f=a(t_f)H$.
 \label{foot:tk}}
 \dis{H\delta t =\kappa\delta {\cal R}=\frac{\kappa}{\sqrt{2\epsilon_H}}\frac{H}{2\pi}(H\Delta t)^{1/2}.\label{eq:tfluc}}
 Here the factor $1/\sqrt{2\epsilon_H}$ comes from the same factor in front of $\varphi$ in the expression for ${\cal R}$ given by \eqref{eq:R}.
 This explicitly shows that the fluctuation in time is accumulated as $t^{1/2}$.

When $H\Delta t \sim {\cal O}(1)$, which corresponds to the early stage of the Hawking radiation in the case of the evaporating black hole, the fluctuation in time $H\delta t \ll 1$ is sufficiently small provided
\dis{\epsilon_H \gg \kappa^2 H^2,}
revealing another tension between the locality and the isometry : whereas we have seen in Sec. \ref{Sec:LGO} that the local gauge invariant operator can be constructed by breaking the isometry, it tells us that the local region   is not well defined if the breaking of the isometry is too small.
In fact, this bound is known as the condition that   eternal inflation does not take place \cite{Olum:2012bn, Matsui:2018bsy, Kinney:2018kew, Rudelius:2019cfh} (see also Sec. 2 of \cite{Seo:2022uaz} for a compact summary).

 On the other hand, the (quasi-)dS space counterpart of the Page time of black hole, $t_{\rm Page}\sim \kappa^{4}M^{3} \sim R^3/\kappa^2$, is the `quantum break time'  given by $t_{\rm QB}\sim H^{-3}\kappa^{-2}=R_{\rm hor}^3/\kappa^2$ \cite{Dvali:2017eba}, the time scale after which quantum fluctuations begin to change the classical trajectory  significantly such that the background is strongly deformed from dS space.
Since $\delta t \propto t^{1/2}/\sqrt{\epsilon_H}$, we expect that the increasing behavior of the fluctuation    $\delta t$ in time can be compensated by the sizeable value of $\epsilon_H$ after $t_{\rm QB}$.
 Indeed, at $t_{\rm QB}$, the fluctuation in time behaves as $H\delta t \sim 1/\sqrt{\epsilon_H} $, which is too large for $\epsilon_H \ll 1$.
 The only way that $H\delta t$ can be sufficiently small is to break the dS isometry strongly  giving $\epsilon_H>1$, which requires that dS space is unstable so rapidly deformed by   the fluctuations.
 The instability of dS space has been conjectured under the name of `dS swampland conjecture', motivated by the difficulty in constructing the string model for the metastable dS vacuum  \cite{Danielsson:2018ztv, Obied:2018sgi, Andriot:2018wzk, Garg:2018reu, Ooguri:2018wrx, Hebecker:2018vxz, Andriot:2018mav} (see also \cite{Lehnert:2025izp} for a recent review focusing on the  cosmological implications). 
In fact, it is closely connected to the second concern in \cite{Antonini:2025sur}, which states that  even if  the timelike isometry is broken by the background describing  the evaporating black hole, the local operators are difficult to be constructed when the isometry breaking effect is parametrically suppressed by $\kappa^2$.

 In order to apply the discussion so far on quasi-dS space  to the issues concerning the island in the evaporating black hole, we first need to make the correspondence between the quantities in quasi-dS space and those in black hole clear.
First, the horizon radius of (quasi-)dS space $R_{\rm hor}=1/H$ plays the same role as that of   black hole $R=2GM$.
This is evident from thermodynamic point of view, where, for example, the temperature of the horizon is given by $H/(2\pi)= 1/(2\pi R_{\rm hor})$ for dS space and $1/(8\pi GM)=1/(\kappa^2 M)=1/(4\pi R)$ for black hole, respectively.
Moreover, just as the timelike isometry breaking in quasi-dS space is parametrized by $\epsilon_H=\dot{R}_{\rm hor}$,  that in the evaporating  black hole also can be parametrized by $\dot{R}$. 
 The explicit expression for $\dot{R}$ can be obtained from the black hole evaporation rate.
   Since $R=2G M$,  the Stefan-Boltzmann law, 
 \dis{-\frac{dM}{dt}\sim A T^4 = \big(16\pi G^2 M^2)\Big(\frac{1}{8\pi G M}\Big)^4 \sim \frac{1}{\kappa^4 M^2},}
 which in fact gives the estimation of the Page time $t_{\rm Page}\sim \kappa^{4}M^{3}$, is rewritten as
 \dis{\dot{R}=-c^2\frac{\kappa^2}{R^2},\quad {\rm or}\quad R=R_0\Big(1-c^2\frac{3\kappa^2 }{ R_0^3}t\Big)^{1/3}, \label{eq:BHcase}}
for some positive constant $c^2$.
Hence, the value of $|\dot{R}|$, the order parameter for  the spontaneous breaking of isometry,    is typically suppressed by $\kappa^2$ in the semiclassical regime $R \gg \kappa$.
Moreover, one also finds that as the black hole evaporates, the black hole radius decreases, resulting in the increase of $\dot{R}$ in time.
Then we expect that after the Page time, $R$ rapidly decreases to $\kappa$, and eventually,   when $R \sim \kappa$,    the timelike isometry of the black hole is strongly broken as $|\dot{R}|$ becomes sizeable,  which is similar situation to $\epsilon_H \sim {\cal O}(1)$ in quasi-dS space.

 Since the island appears in the black hole interior after $t_{\rm Page}$, in order for the island can be reliably constructed,  the timelike isometry must be strongly broken  after $t_{\rm Page}$ such that the  fluctuation of spacetime point is sufficiently suppressed by $|\dot{R}|\gtrsim {\cal O}(1)$ against its behavior  $\sim t^{1/2}$.
 However, as will be shown,  it is difficult to be achieved if we assume that spontaneous breaking of the timelike isometry in the evaporating black hole is described in the similar manner to that in quasi-dS space.
 This assumption is supported by our expectation that in both quasi-dS space and evaporating black hole, the spontaneous breaking of the timelike isometry leads to the local gauge invariant operator  made up of the metric fluctuation (which can be identified with the fluctuation in time) and some matter  field one (whose classical solution depends on time so breaks the timelike isometry).
 Since the black hole radius $R$, or equivalently, the inverse of temperature, is  interpreted as a typical time scale of the black hole evaporation, we may identify $\kappa \zeta$, the metric fluctuation in the direction of time translation, with $\delta t/R$, just like $\kappa \zeta \sim H\delta t$ in quasi-dS space. 
 This is promoted to the local gauge invariant operator by absorbing the matter field fluctuation, say, the scalar fluctuation $\varphi$, in the form of $\zeta-\varphi/(\kappa \dot{\phi}_0 R)$, which is nothing more than the form of the curvature perturbation in quasi-dS space given by \eqref{eq:R}.
 Moreover, just like \eqref{eq:epsilonH}, the nonzero $\dot{R}$ and $\dot{\phi}_0$ parametrizing the spontaneous breaking of the timelike isometry are connected through the equations of motion.
 Since both quantities vanish when the isometry is restored,  $\dot{\phi}_0$ would be  expressed as a some positive power of $\dot{R}$.
Finally, as $R$ decreases due to the black hole evaporation, the longest wavelength $\sim R$ relevant to dynamics of the black hole interior also decreases.
Then as time goes on, more and more modes  with wavelength longer than $R$ will be  frozen out. 
They are accumulated to give rise to  the fluctuation of the spacetime point, which increases as $\sim t^{1/2}$ : this indeed is a typical feature of the statistical fluctuation for the system resembling the random walk.

Discussion so far suggests that the fluctuation in time can be constructed in the same manner as \eqref{eq:tfluc} hence given by
\dis{  \frac{\delta t}{R}\sim \frac{1}{|\dot{R}|^{n/2}}\frac{\kappa}{R}\Big(\frac{\Delta t}{R}\Big)^{1/2},\label{eq:noisland}}
for some positive $n$.
Combining this with \eqref{eq:BHcase}, we obtain
\dis{\frac{\delta t}{R} \sim \kappa^{1-n}R^{n-\frac{3}{2}}t^{1/2}=\kappa^{1-n}R_0^{n-\frac{3}{2}}\Big(1-c^2\frac{3\kappa^2}{R_0^3}t\Big)^{\frac13 \big(n-\frac{3}{2}\big) }t^{1/2}.}
Requiring that $\delta t$ is suppressed by the positive power of $\kappa$ for $t/R\ll 1$, $n$ must be smaller than $1$.
However, if $n$ is smaller than $3/2$, $\delta t/R$ increases monotonically in time, which means that even though $|\dot{R}|$ increases in time, it is not sufficient to compensate the $t^{1/2}$ behavior of $\delta t/R$.
 Therefore, there is no appropriate value of $n$ which suppresses the rate $\delta t/R$ at early and late stages of the Hawking radiation simultaneously.
 In fact, for $1/2<n<1$, $\delta t$ increases until $t_{\rm max} =R_0^3/[2(n+1) c^2 \kappa^2]$ (which is of the same order as $t_{\rm Page}$) and then decreases rapidly.
 However, it is not enough to change the increasing behavior of the ratio $\delta t/R$ against decreasing value of $R$ at late time, as the above expression tells us.

 This may be resolved, for example, if the expression for   $|\dot{R}|$ or $\delta t/R$ is modified  at late time.
 Indeed, if we set $\dot{R}=-c^2(\kappa/R)^m$, the black hole radius becomes
 \dis{R=R_0 \Big(1-c^2(m+1)\frac{\kappa^m}{R_0^{m+1}}t\Big)^{\frac{1}{m+1}},}
 from which we obtain
 \dis{\frac{\delta t}{R} \sim \kappa^{1-\frac{n m}{2}}R_0^{\frac12(nm-3)}\Big(1-c^2(m+1)\frac{\kappa^m}{R_0^{m+1}}t\Big)^{\frac{1}{2(m+1)}(nm-3)}.} 
 Then even if $\delta t/R$ is suppressed at   early stage $t/R\ll 1$ when $m=2$ hence $n$ is smaller than $1$, it can decrease  in time at late stage if at least one of $n$ and $m$ changes such that  $nm>3$ is satisfied.
 For example, suppose the black hole at late stage can see the hidden extra dimensions.
 Since  in $d$ spacetime dimensions the black hole radius and the mass are related by 
 \dis{R^{d-3}=\frac{16\pi GM}{(d-2)\Omega_{d-2}}}
while the temperature is given by $(d-3)/(4\pi R)$ \cite{Emparan:2008eg}, the Stefan-Boltzmann law $-\dot{M}=A T^d$ is rewritten as
\dis{\dot{R}=-c^2\frac{\kappa^{d-2}}{R^{d-2}}.} 
That is, as $d$ becomes larger than $4$, $m=d-2$ also increases as well.
Therefore, so far as the expression \eqref{eq:noisland} is not changed in higher dimensional spacetime except for the increase in the value of $n$, the fluctuation $\delta t/R$ decreases at late stage of the black hole evaporation, which allows the reliable construction of the local region at late state of the black hole evaporation.

 But even in this case, there is another challenge for construction of the island described by local operators only, which comes from the fact that the island solution relies on the nonlocal nature of quantum gravity.
More concretely, in order for the island to resolve the black hole information paradox, operators on the island must contain information on the Hawking radiation far away outside the black hole horizon.
But if the operators on the island are local, it is not clear at present that how such nonlocal feature can be realized.
Regarding this, the recent studies \cite{Geng:2025byh, Geng:2025rov} claim that when diffeomorphism (rather than isometry) is broken by the boundary condition (where the boundary refers to that between the bulk and the bath discussed in Introduction), the massive graviton encodes information on the heat bath.
Then the compensating vector field plays the role of the wormhole operator connecting the island and the bath.
Whether the similar mechanism can be realized when the isometry is spontaneously broken (without breaking diffeomorphism) is a subject of future investigation.


 \section{Conclusion}
 
 In this article, we have investigated the possibility of locality without breaking diffeomorphism invariance.
 Here by locality we mean the commutativity of operators  on different space-like separated region.
  When the isometry is spontaneously broken, the local gauge invariant operators can be constructed without breaking the diffeomorphism invariance.
 This can be achieved by the St\"uckelberg mechanism, where the fluctuation of the metric in the direction of the broken isometry transformation is combined with that of the matter field whose classical solution breaks the isometry to become physical.
 These  operators  also can be interpreted as the gauge invariant promotion of the local operators by introducing the `clock' and the `rod', which distinguish some particular time or spatial slice from another. 
 Indeed, the concept of the clock and the rod has been employed to describe the thermodynamic properties of quantum gravity in terms of the von Neumann algebra \cite{Chandrasekaran:2022cip, Seo:2022pqj,  Gomez:2023wrq, Kudler-Flam:2024psh, Geng:2025bcb} (for reviews on von Neumann algebra in quantum gravity, see, e.g., \cite{Witten:2018zxz, Liu:2025krl}).

 The construction of the local gauge invariant operator described above shows the tension between the locality and the isometry.
 Such a tension also can be found when we consider the effects of the accumulated fluctuation of the spacetime point generated by the frozen modes in the presence of the horizon.
 Since this fluctuation increases as $t^{1/2}$, the local region like the island may not be well constructed even if the local gauge invariant operators exist.
 In quasi-dS space describing the inflationary cosmology, the fluctuation of the spacetime point can be suppressed when the background is significantly deformed from dS space such that the timelike isometry is strongly broken.
 The structure of the local gauge invariant operator suggests that the similar situation can be found in the black hole interior after the Page time.
 In this case, the strong deformation of the background geometry   may not be enough to suppress the fluctuation   against its increasing behavior $\sim t^{1/2}$, unless the black hole can see the extra dimensions at late stage of the evaporation.
 This condition on the accumulated fluctuation provides a stringent constraint for the appropriate implementation of the island, in addition to the existence of the local operator.
 
  We finally note that even though the spontaneous breaking of the isometry reconciles the locality and the diffeomorphism invariance,  it may give rise to another issue.
  For example, the existence of the timelike Killing vector is an essential ingredient to define the positive and the negative modes appropriately, which is the basis of the positive energy theorem \cite{Witten:1981mf, Parker:1981uy}.
  However, if the timelike isometry is spontaneously broken, such a separation of modes may not be well defined. 
  While challenging, clarifying this issue can be an important step in finding out the bridge between the local quantum field theory and quantum gravity.

%


\appendix



\renewcommand{\theequation}{\Alph{section}.\arabic{equation}}


\subsection*{Acknowledgements}


\end{document}